\documentclass[useAMS,usenatbib]{mn2e}

\usepackage{graphicx}
\usepackage{amssymb}
\usepackage{amsmath}

\newcommand{\feh}{\mbox{[Fe/H]}}
\newcommand{\teff}{\mbox{$T_{\rm eff}$}}
\newcommand{\logg}{\mbox{$\log g_*$}}
\newcommand{\vsini}{\mbox{$v \sin I$}}
\newcommand{\mictrb}{\mbox{$\xi_{\rm t}$}}
\newcommand{\mactrb}{\mbox{$v_{\rm mac}$}}

\newcommand{\kms}{\mbox{km\,s$^{-1}$}}
\newcommand{\ms}{\mbox{m\,s$^{-1}$}}

\newcommand{\halpha}{\mbox{$H\alpha$}}
\newcommand{\mplanet}{\mbox{$M_{\rm pl}$}}
\newcommand{\rplanet}{\mbox{$R_{\rm pl}$}}
\newcommand{\densplanet}{\mbox{$\rho_{\rm pl}$}}
\newcommand{\mjup}{\mbox{$M_{\rm Jup}$}}
\newcommand{\rjup}{\mbox{$R_{\rm Jup}$}}
\newcommand{\densjup}{\mbox{$\rho_{\rm Jup}$}}
\newcommand{\mstar}{\mbox{$M_*$}}
\newcommand{\rstar}{\mbox{$R_*$}}
\newcommand{\densstar}{\mbox{$\rho_*$}}

\newcommand{\msol}{\mbox{$M_\odot$}}
\newcommand{\rsol}{\mbox{$R_\odot$}}
\newcommand{\denssol}{\mbox{$\rho_\odot$}}

\def\secos{$\sqrt{e} \cos \omega$}
\def\sesin{$\sqrt{e} \sin \omega$}

\def\teql{$T_{\rm eql}$}

\def\chisq{$\chi^2$}

\mathchardef\mhyphen="2D
\def\PLS{$\mathbb{P}_{\rm LS}$}

\newcommand{\aap}{A\&A}
\newcommand{\apj}{ApJ}
\newcommand{\apjl}{ApJ}
\newcommand{\mnras}{MNRAS}
\newcommand{\apjs}{ApJS}

\newcommand{\pasp}{PASP}
\newcommand{\aapr}{A\&A~Rev.}
\newcommand{\aaps}{A\&AS}
\newcommand{\aj}{AJ}

\title[WASP-44b, WASP-45b and WASP-46b]
{WASP-44b, WASP-45b and WASP-46b: three short-period, transiting extrasolar 
planets}
\author[D.~R.~Anderson et al.]
{D.~R.~Anderson,$^{1}$\thanks{dra@astro.keele.ac.uk}
A.~Collier~Cameron,$^{2}$ 
M.~Gillon,$^{3}$ 
C.~Hellier,$^1$ 
E.~Jehin,$^3$
M.~Lendl,$^4$ 
\newauthor
P.~F.~L.~Maxted,$^1$ 
D.~Queloz,$^4$ 
B.~Smalley,$^1$ 
A.~M.~S.~Smith,$^1$ 
A.~H.~M.~J.~Triaud,$^4$
\newauthor
R.~G.~West,$^5$ 
F.~Pepe,$^4$
D.~Pollacco,$^6$
D.~S\'egransan,$^4$
I.~Todd$^6$ and 
S.~Udry$^4$\\
$^1$Astrophysics Group, Keele University, Staffordshire ST5 5BG, UK\\
$^2$SUPA, School of Physics and Astronomy, University of St. Andrews, 
North Haugh, Fife, KY16 9SS, UK\\
$^3$Institut d'Astrophysique et de G\'eophysique,  Universit\'e de 
Li\`ege,  All\'ee du 6 Ao\^ut, 17,  Bat.  B5C, Li\`ege 1, Belgium\\
$^4$Observatoire de Gen\`eve, Universit\'e de Gen\`eve, 51 Chemin 
des Maillettes, 1290 Sauverny, Switzerland\\
$^5$Department of Physics and Astronomy, University of Leicester, 
Leicester, LE1 7RH, UK\\
$^6$Astrophysics Research Centre, School of Mathematics \& Physics,
 Queen's University, University Road, Belfast, BT7 1NN, UK}

\begin{document}

\date{Accepted 2012 January 21. Received 2012 January 21; in original form 2011 May 15}

\pagerange{\pageref{firstpage}--\pageref{lastpage}} \pubyear{2011}

\maketitle

\label{firstpage}

\begin{abstract}
We report the discovery of three extrasolar planets that transit their 
moderately bright ($m_V$ = 12--13) host stars. 
WASP-44b is a 0.89-\mjup\ planet in a 2.42-day orbit around a G8V star. 
WASP-45b is a 1.03-\mjup\ planet which passes in front of the limb of its K2V 
host star every 3.13 days. 
Weak Ca {\sc ii} H+K emission seen in the spectra of WASP-45 suggests the star 
is chromospherically active. 
WASP-46b is a 2.10-\mjup\ planet in a 1.43-day orbit around a G6V star. 
Rotational modulation of the light curves of WASP-46 and weak Ca {\sc ii} H+K 
emission in its spectra show the star to be photospherically and 
chromospherically active. 

We imposed circular orbits in our analyses as the radial velocity data are 
consistent with (near-)circular orbits, as could be expected from both empirical 
and tidal-theory perspectives for such short-period, $\sim$Jupiter-mass 
planets.
We discuss the impact of fitting for eccentric orbits for such planets 
when not supported by the data. 
The derived planetary and stellar radii depend on the fitted eccentricity and 
these parameters inform intense theoretical efforts concerning 
tidal circularisation and heating, bulk planetary composition and the 
observed systematic errors in planetary and stellar radii. 
As such, we recommend exercising caution in fitting the orbits of short period, 
$\sim$Jupiter-mass planets with an eccentric model when there is no 
evidence of non-circularity. 
\end{abstract}


\section{Introduction}

The ensemble of well-characterised transiting extrasolar planets is growing 
at pace, with well over one hundred known to date.
It is important to determine the system parameters accurately so that the 
inferrences based on them are reliable. 
For example, to determine the bulk composition of a planet it is necessary to 
measure accurately its radius \citep[e.g.][]{2007ApJ...659.1661F}. 
Many short-period, giant planets 
(e.g. WASP-17b, \citealp{2010ApJ...709..159A}, \citealp{2011MNRAS.416.2108A}) 
are larger 
than predicted by standard cooling theory of irradiated, gas-giant planets 
\citep[e.g.][]{2007ApJ...659.1661F}. 
One potential explanation is that energy from the tidal circularisation 
of eccentric orbits was dissipated within the planets' interiors, causing them 
to bloat \citep[e.g.][]{2001ApJ...548..466B}. 
To evaluate the likelihood that a planet was inflated by such tidal heating, it 
is necessary to have an accurate determination of both its radius and its 
orbital eccentricity \citep[e.g.][]{2011ApJ...727...75I}. 

A planet's orbital eccentricity can be determined by measuring the radial 
motion of its host star around its orbit 
\citep[e.g.][]{2010A&A...517L...1Q}, 
or by observing occultations of the planet by its host star 
\citep[e.g.][]{2011MNRAS.416.2108A}, 
or from a combination of the two. 
By combining this eccentricity measurement with high-quality transit light 
curves, we can measure a star's density \citep{2003ApJ...585.1038S}. 
The stellar mass can be estimated using stellar evolution models 
\citep[e.g.][]{2004ApJS..155..667D}
or mass-calibration laws \citep[e.g.][]{2010A&ARv..18...67T}, 
and the stellar radius follows. 
This, combined with the ratio of the radii derived from the transit depth, gives 
the planet radius. 

In papers announcing new planet discoveries, eccentricity is often poorly 
determined, thus the planet radius can be uncertain. 
Despite this, there are theoretical \citep{1966Icar....5..375G} and empirical 
\citep{2011MNRAS.tmp..378P} 
reasons to expect short-period ($\lesssim 4$ d), Jupiter-mass ($\approx$ 0.5--2 
\mjup) planets, often referred to as {\it hot Jupiters}, to be in circular 
orbits. 
Thus it is reasonable, in the absence of evidence to the contrary, to assume 
that a newly discovered hot Jupiter is in a circular orbit. 
In so doing, the derived stellar and planetary dimensions will be, on the whole, 
accurate. 

In this paper we present three new hot Jupiters, WASP-44b, WASP-45b and 
WASP-46b, and discuss the effects of fitting an eccentric orbit model to a 
hot Jupiter system in the absence of evidence for non-circularity.

\section{Observations}


\subsection{Discovery photometry}
\label{sec:disc-phot}
The WASP (Wide Area Search for Planets) photometric survey 
\citep{2006PASP..118.1407P} employs two eight-camera arrays, each with a 
field of view of 450 deg$^2$, to monitor bright stars ($m_{\rm V}$ = 8--15). 
Each array observes up to eight pointings per night with a cadence of 
5--10 min, and each pointing is followed for around five months at a time. 
The WASP-South station \citep{2011EPJWC..1101004H} is hosted by 
the South African Astronomical Observatory and the SuperWASP-North station 
\citep{2011EPJWC..1101003F} is hosted by 
the Observatorio del Roque de Los Muchachos on La Palma. 
The WASP data were processed and searched for transit signals as described in 
\citet{2006MNRAS.373..799C} and the candidate selection process is described in 
\citet{2007MNRAS.380.1230C}.

\begin{figure}
\centering                     
\includegraphics[width=8.4cm]{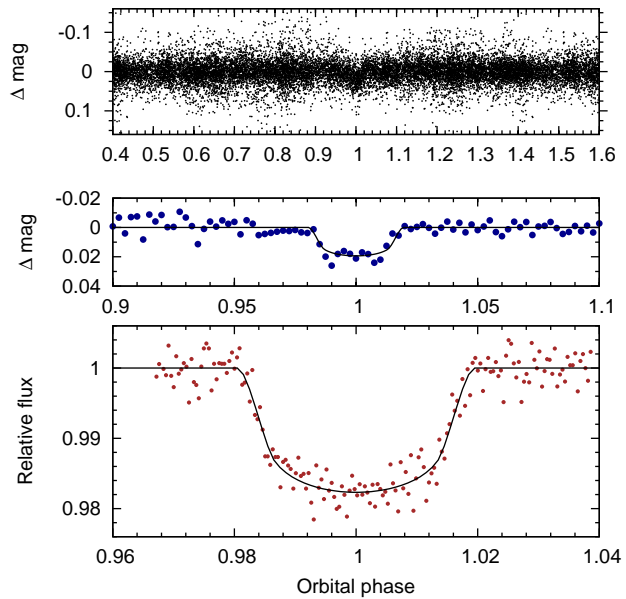}
\caption{Photometry of WASP-44, with the best-fitting transit model 
superimposed. 
{\bfseries Top:} WASP discovery light curve, folded on the ephemeris 
of Table~\ref{tab:mcmc}. 
{\bfseries Middle:} WASP data around the transit, binned in time with a bin 
width of $\sim$9 min.
{\bfseries Bottom:} High-precision transit light curve from Euler.}
\label{fig:w44-phot}
\end{figure}

WASP-44 is a $m_{\rm V}$ = 12.9, G8V star located in the constellation Cetus. 
A transit search of the light curve obtained by WASP-South from 2009 July to 
November found a weak, 2.42-d periodicity.  Further observations in 2008 and 
2009 with both WASP instruments led to a total of 15\,755 photometric 
measurements (Figure~\ref{fig:w44-phot}).
WASP-44 was also observed by a SuperWASP-North camera during 2010 August to 
November.
The resulting light curve comprises 6\,000 measurements and became 
available during the preparation of this paper. As such, it was not used in 
determination of the system paramters (Section~\ref{sec:syspar}), but was used 
in the search for rotational modulation (Section~\ref{sec:rot}).

\begin{figure}
\centering                     
\includegraphics[width=8.4cm]{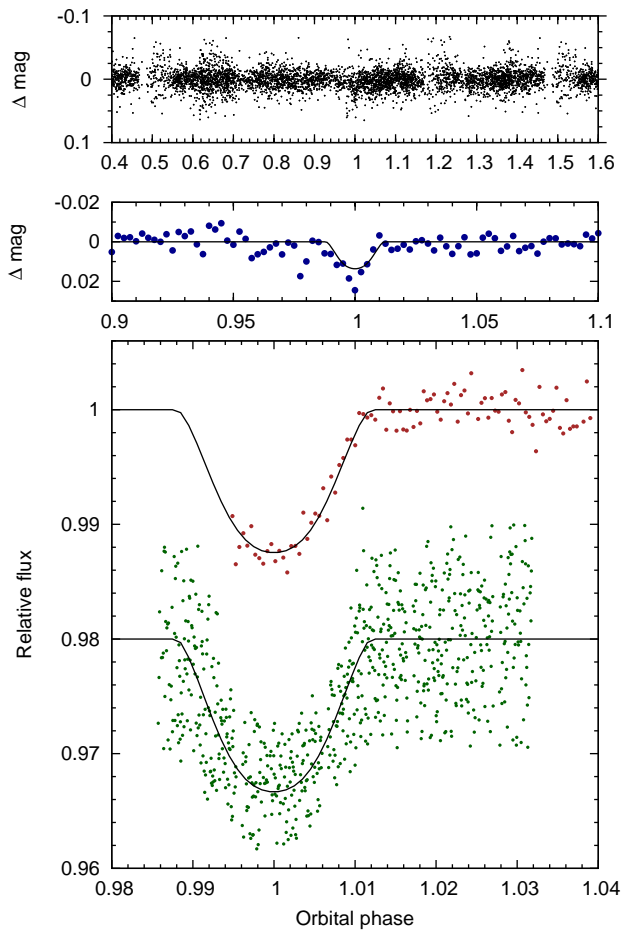}
\caption{Photometry of WASP-45, with the best-fitting transit model 
superimposed. 
{\bfseries Top:} WASP discovery light curve, folded on the ephemeris 
of Table~\ref{tab:mcmc}. 
{\bfseries Middle:} WASP data around the transit, binned in time with a bin 
width of $\sim$11 min.
{\bfseries Bottom:} High-precision transit light curves from Euler (upper) and 
TRAPPIST (lower).}
\label{fig:w45-phot}
\end{figure}

WASP-45 is a $m_{\rm V}$ = 12.0, K2V star located in the consellation Sculptor 
that was observed by WASP-South during May to November of 2006 and 2007. 
A transit search of the resulting light curve, which comprises 11\,007 
photometric measurements, found a strong 3.13-d periodicity 
(Figure~\ref{fig:w45-phot}). 

\begin{figure}
\centering                     
\includegraphics[width=8.4cm]{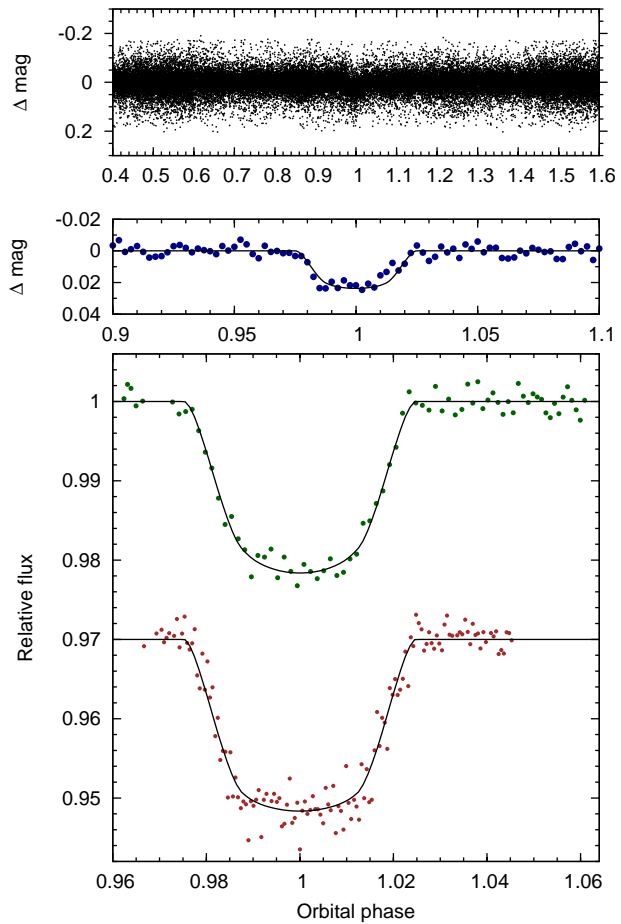}
\caption{Photometry of WASP-46, with the best-fitting transit model 
superimposed. 
{\bfseries Top:} WASP discovery light curve, folded on the ephemeris 
of Table~\ref{tab:mcmc}. 
{\bfseries Middle:} WASP data around the transit, binned in time with a bin 
width of $\sim$5 min.
{\bfseries Bottom:} High-precision transit light curves from TRAPPIST (upper) and 
Euler (lower).}
\label{fig:w46-phot}
\end{figure}

WASP-46 is a $m_{\rm V}$ = 12.9, G6V star located in the constellation Indus 
that was observed by two WASP-South cameras during May to October of 2008 and 
2009. 
A transit search of the resulting light curve, which comprises 41\,961 
photometric measurements, found a strong 1.43-d periodicity 
(Figure~\ref{fig:w46-phot}). 
WASP-46 was also observed by two WASP-South cameras during 2010 August to 
November.  
The resulting light curve comprises 19\,000 measurements and became 
available during the preparation of this paper. 
As such, it was not used in the determination of the system paramters 
(Section~\ref{sec:syspar}), but was used in the search for rotational 
modulation (Section~\ref{sec:rot}).

WASP-46 is blended in the WASP images with a $m_{\rm V}$ = 14.8 star at a 
separation of 17.4\arcsec\ (the plate scale is 13.7\arcsec pixel$^{-1}$ and 
the photometry aperture has a radius of 3.5 pixels). 
We corrected the WASP photometry for this contamination to prevent dilution of 
the transit. 

\subsection{Spectroscopic follow-up}

\begin{figure}
\centering                     
\includegraphics[width=8.4cm]{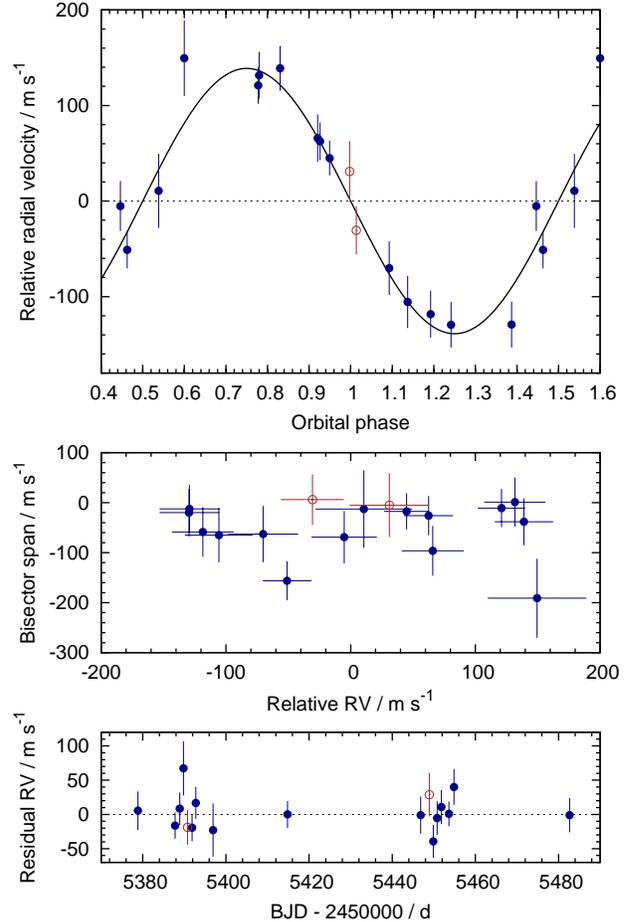}
\caption{{\bfseries Top:} CORALIE RVs of WASP-44 with the best-fitting circular 
Keplerian orbit superimposed (rms=25.0\,\ms, \chisq=10.9).
The RVs represented by open, red circles were excluded from the analysis as they 
were taken during transit. 
{\bfseries Middle:} Bisector span with respect to RV. 
We adopted uncertainties on the bisector spans twice the size of those on the 
RVs.
{\bfseries Bottom:} Residuals of the RVs about the fit as a function of time.}
\label{fig:w44-rv}
\end{figure}

\begin{figure}
\centering                     
\includegraphics[width=8.4cm]{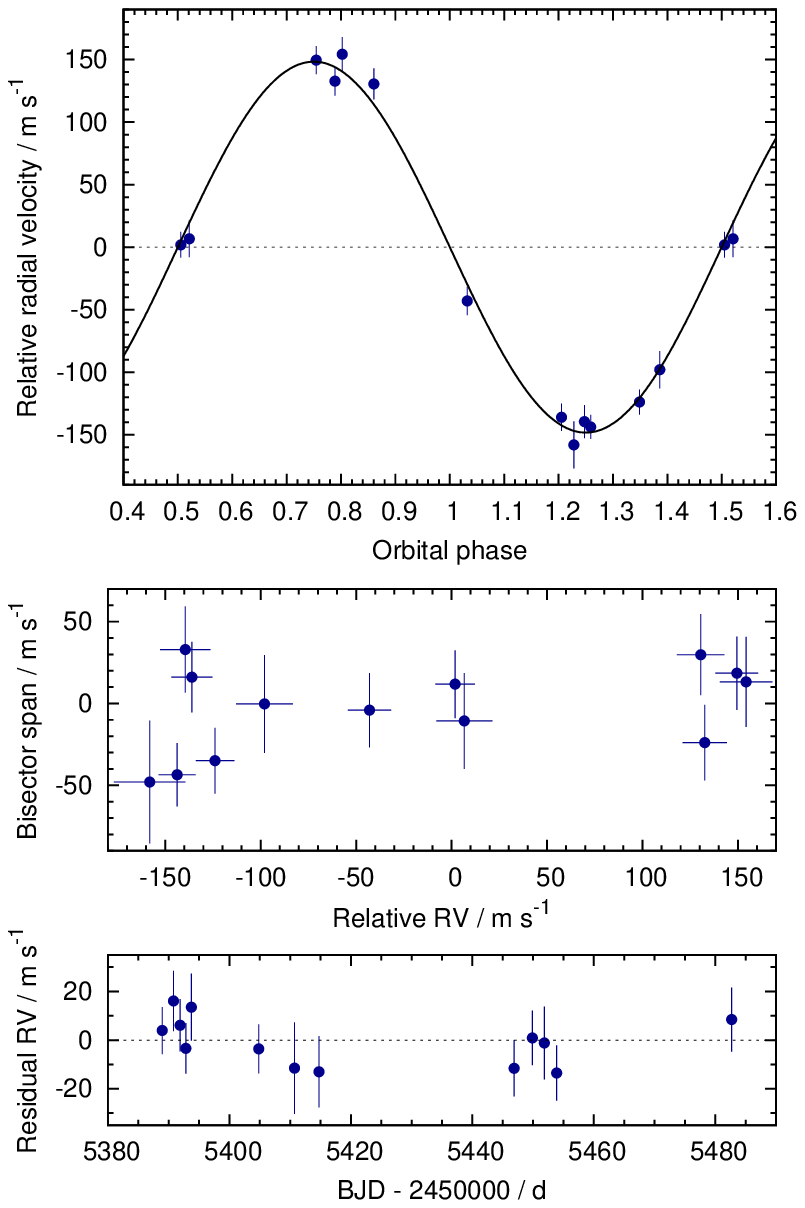}
\caption{{\bfseries Top:} CORALIE RVs of WASP-45 with the best-fitting circular 
Keplerian orbit superimposed (rms=9.7\,\ms, \chisq=7.4).
{\bfseries Middle:} Bisector span with respect to RV. 
We adopted uncertainties on the bisector spans twice the size of those on the 
RVs.
{\bfseries Bottom:} Residuals of the RVs about the fit as a function of time.}
\label{fig:w45-rv}
\end{figure}

\begin{figure}
\centering                     
\includegraphics[width=8.4cm]{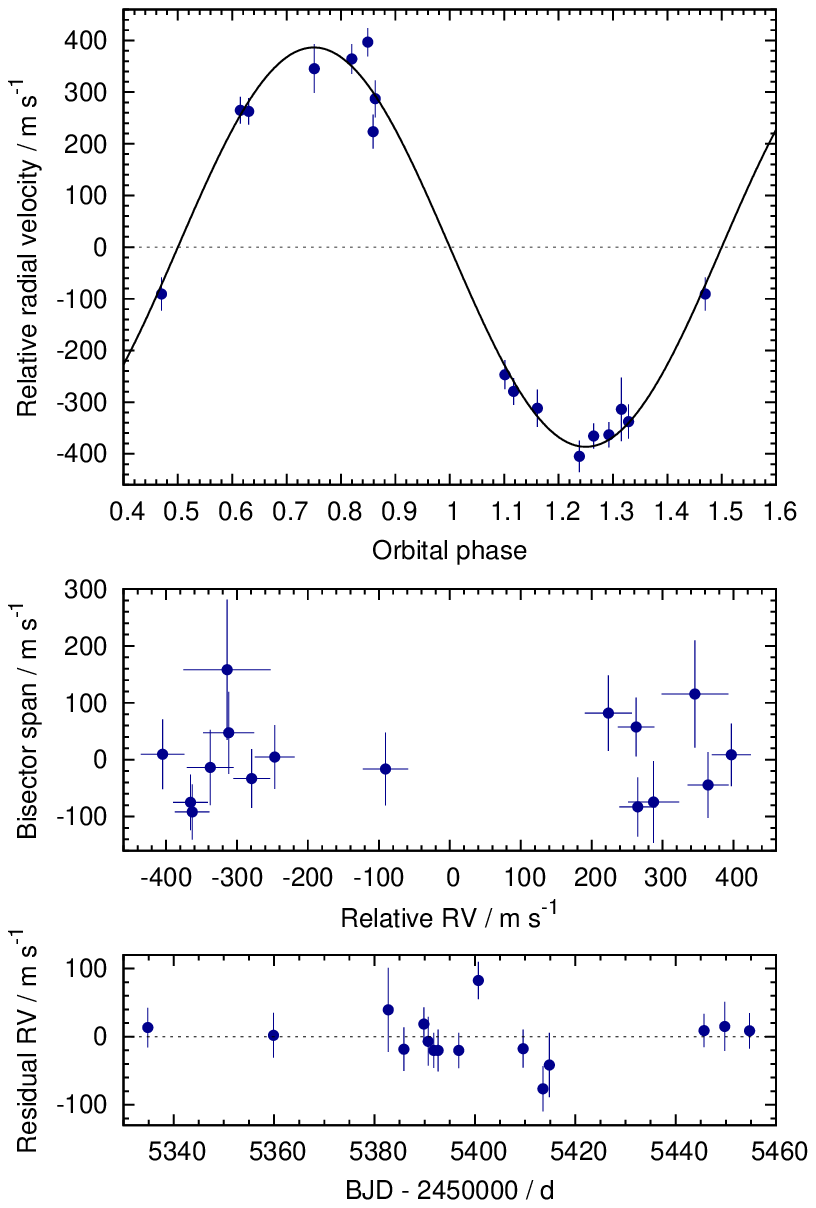}
\caption{{\bfseries Top:} CORALIE RVs of WASP-46 with the best-fitting circular 
Keplerian orbit superimposed (rms=34.2\,\ms, \chisq=19.2).
{\bfseries Middle:} Bisector span with respect to RV. 
We adopted uncertainties on the bisector spans twice the size of those on the 
RVs.
{\bfseries Bottom:} Residuals of the RVs about the fit as a function of time.}
\label{fig:w46-rv}
\end{figure}

In 2010 we used the CORALIE spectrograph mounted on the 1.2-m Euler-Swiss 
telescope \citep{1996A&AS..119..373B,2000A&A...354...99Q} to obtain spectra of 
the three target stars. 
We obtained 17 spectra of WASP-44, 13 spectra of WASP-45 and 16 spectra of 
WASP-46.
Radial-velocity (RV) measurements were computed by weighted cross-correlation 
\citep{1996A&AS..119..373B,2005Msngr.120...22P} with a numerical G2-spectral 
template for both WASP-44 and WASP-46, and with a K5-spectral template for 
WASP-45.
RV variations were detected with the same periods found from the WASP photometry 
and with semi-amplitudes consistent with planetary-mass companions. 
The RV measurements are listed in Table~\ref{tab:rv} and are plotted separately 
for each system in Figures~\ref{fig:w44-rv}, \ref{fig:w45-rv} and 
\ref{fig:w46-rv}.
We excluded two RVs of WASP-44 from the analysis as the spectra were taken 
during transit and we do not fit for the Rossiter-McLaughlin effect 
\citep[e.g.][]{2011MNRAS.414.3023S}.

For each system we tested the hypothesis that the RV variations are due to 
spectral-line distortions caused by a blended eclipsing binary or starspots  
by performing a line-bisector analysis \citep{2001A&A...379..279Q} of the 
CORALIE cross-correlation functions. 
The lack of correlation between bisector span and RV (Figures~\ref{fig:w44-rv}, 
\ref{fig:w45-rv} and \ref{fig:w46-rv}) supports our conclusion that the periodic 
dimming and RV variation of each system are caused by a transiting planet.

\begin{table} 
\caption{Radial velocity measurements} 
\label{tab:rv} 
\begin{tabular*}{0.5\textwidth}{@{\extracolsep{\fill}}lllll} 
\hline 
Star & BJD (UTC) & RV & $\sigma$$_{\rm RV}$ & BS\\ 
 & (day) & (km s$^{-1}$) & (km s$^{-1}$) & (km s$^{-1}$)\\ 
\hline
WASP-44 & 2\,455\,378.8533 & $-$4.114 & ~~\,0.028 & $-$0.063\\
WASP-44 & 2\,455\,387.7850 & $-$3.923 & ~~\,0.019 & $-$0.011\\
WASP-44 & 2\,455\,388.9089 & $-$4.173 & ~~\,0.024 & $-$0.020\\
WASP-44 & 2\,455\,389.7778 & $-$3.894 & ~~\,0.040 & $-$0.191\\
WASP-44 & 2\,455\,390.7802$^{\rm a}$ & $-$4.074 & ~~\,0.025 & ~~\,0.006\\
WASP-44 & 2\,455\,391.8691 & $-$4.095 & ~~\,0.020 & $-$0.156\\
WASP-44 & 2\,455\,392.7604 & $-$3.905 & ~~\,0.023 & $-$0.038\\
WASP-44 & 2\,455\,396.8995 & $-$4.033 & ~~\,0.039 & $-$0.013\\
WASP-44 & 2\,455\,414.8075 & $-$3.981 & ~~\,0.020 & $-$0.026\\
WASP-44 & 2\,455\,446.8271 & $-$4.149 & ~~\,0.027 & $-$0.065\\
WASP-44 & 2\,455\,448.9128$^{\rm a}$ & $-$4.013 & ~~\,0.032 & $-$0.005\\
WASP-44 & 2\,455\,449.8568 & $-$4.173 & ~~\,0.024 & $-$0.012\\
WASP-44 & 2\,455\,450.8095 & $-$3.912 & ~~\,0.025 & ~~\,0.001\\
WASP-44 & 2\,455\,451.8089 & $-$4.162 & ~~\,0.025 & $-$0.059\\
WASP-44 & 2\,455\,453.6439 & $-$3.999 & ~~\,0.018 & $-$0.017\\
WASP-44 & 2\,455\,454.8482 & $-$4.049 & ~~\,0.026 & $-$0.069\\
WASP-44 & 2\,455\,482.6600 & $-$3.978 & ~~\,0.025 & $-$0.096\\

WASP-45 & 2\,455\,388.9353 & ~~\,4.406 & ~~\,0.010 & $-$0.044\\
WASP-45 & 2\,455\,390.8149 & ~~\,4.680 & ~~\,0.012 & ~~\,0.030\\
WASP-45 & 2\,455\,391.8937 & ~~\,4.414 & ~~\,0.011 & ~~\,0.016\\
WASP-45 & 2\,455\,392.8314 & ~~\,4.552 & ~~\,0.010 & ~~\,0.012\\
WASP-45 & 2\,455\,393.7593 & ~~\,4.704 & ~~\,0.014 & ~~\,0.013\\
WASP-45 & 2\,455\,404.8466 & ~~\,4.426 & ~~\,0.010 & $-$0.035\\
WASP-45 & 2\,455\,410.7215 & ~~\,4.392 & ~~\,0.019 & $-$0.048\\
WASP-45 & 2\,455\,414.7627 & ~~\,4.557 & ~~\,0.015 & $-$0.011\\
WASP-45 & 2\,455\,446.8614 & ~~\,4.683 & ~~\,0.012 & $-$0.024\\
WASP-45 & 2\,455\,449.8794 & ~~\,4.699 & ~~\,0.011 & ~~\,0.018\\
WASP-45 & 2\,455\,451.8550 & ~~\,4.452 & ~~\,0.015 & $-$0.000\\
WASP-45 & 2\,455\,453.8741 & ~~\,4.507 & ~~\,0.011 & $-$0.004\\
WASP-45 & 2\,455\,482.6826 & ~~\,4.410 & ~~\,0.013 & ~~\,0.033\\

WASP-46 & 2\,455\,334.8428 & $-$3.414 & ~~\,0.029 & $-$0.044\\
WASP-46 & 2\,455\,359.8870 & $-$4.116 & ~~\,0.033 & $-$0.014\\
WASP-46 & 2\,455\,382.7544 & $-$4.092 & ~~\,0.062 & ~~\,0.158\\
WASP-46 & 2\,455\,385.8357 & $-$3.869 & ~~\,0.032 & $-$0.016\\
WASP-46 & 2\,455\,389.8330 & $-$4.144 & ~~\,0.024 & $-$0.075\\
WASP-46 & 2\,455\,390.6889 & $-$3.491 & ~~\,0.036 & $-$0.074\\
WASP-46 & 2\,455\,391.7867 & $-$3.516 & ~~\,0.026 & ~~\,0.057\\
WASP-46 & 2\,455\,392.6562 & $-$4.183 & ~~\,0.031 & ~~\,0.010\\
WASP-46 & 2\,455\,396.7744 & $-$4.058 & ~~\,0.026 & $-$0.033\\
WASP-46 & 2\,455\,400.6822 & $-$3.382 & ~~\,0.027 & ~~\,0.009\\
WASP-46 & 2\,455\,409.6251 & $-$4.025 & ~~\,0.028 & ~~\,0.005\\
WASP-46 & 2\,455\,413.5695 & $-$3.555 & ~~\,0.033 & ~~\,0.082\\
WASP-46 & 2\,455\,414.8452 & $-$3.433 & ~~\,0.047 & ~~\,0.115\\
WASP-46 & 2\,455\,445.6578 & $-$4.141 & ~~\,0.024 & $-$0.092\\
WASP-46 & 2\,455\,449.7606 & $-$4.090 & ~~\,0.036 & ~~\,0.047\\
WASP-46 & 2\,455\,454.7010 & $-$3.513 & ~~\,0.026 & $-$0.083\\
\hline
\end{tabular*}
$^{\rm a}$ These two RVs were excluded from the analysis as they were measured 
during transit.
\end{table}

\subsection{Follow-up photometry}

We obtained high signal-to-noise (S/N) transit photometry to refine the 
parameters of each system. 
We used the 1.2-m Euler-Swiss telescope located at ESO La Silla 
Observatory, Chile to observe one transit each of WASP-44b, WASP-45b and 
WASP-46b.
Observations were made through a Gunn $r$ filter and Euler's absolute tracking 
mode was used to keep the stars on the same location on the chip. This is done 
by calculating an astrometric solution for each science frame and adjusting the 
telescope pointing between exposures. After correcting the 
images for bias, overscan and flat-field variations, we extracted aperture 
photometry for the bright stars in the field. We selected the aperture radius 
and set of reference stars that resulted in the smallest light curve residuals. 

We used TRAPPIST\footnote{TRAnsiting Planets and PlanetesImals Small Telescope; 
http://arachnos.astro.ulg.ac.be/Sci/Trappist}, a 60 cm robotic telescope also 
located at ESO La Silla \citep{2011EPJWC..1106002G} to observe one 
transit of WASP-45b and one transit of WASP-46b. 
Both transits were observed through a special $I$+$z$ filter, in the 2 
MHz read-out mode, and with 1$\times$1 binning, resulting in a typical combined 
readout and overhead time of 8 s and a readout noise of 13.5 $e^{-}$. The 
`software guiding' system of TRAPPIST kept the stars at the same positions on 
the chip to within a few pixels over the course of the observations 
(Gillon et al., in preparation).

\subsubsection{WASP-44b}

A transit of WASP-44b was observed on 2010 September 14 by Euler. We 
observed for a total of 4.2 hours, from 46 min before the start of the transit 
to 65 min after it ended. The conditions were clear, airmass ranged from 
1.04 to 1.44 and no defocus was applied. 
The WASP-44b transit light curve is shown in Figure~\ref{fig:w44-phot} and the 
data are given in Table~\ref{tab:phot}.

\subsubsection{WASP-45b}

A partial transit of WASP-45b was observed on 2010 September 11 by Euler. 
We observed for a total of 3.3 hours, from part-way through the transit 
to 2 hours after it ended. The seeing was poor (1--2\arcsec) and 
airmass ranged from 1.01 to 1.22. We applied a defocus of 0.1 mm to increase the 
duty cycle and to minimise pixel-to-pixel effects. 

A transit of WASP-45b was observed on 2010 December 17 by TRAPPIST. 
Observations lasted from 0h15 to 3h40 UT, during which the airmass increased 
from 1.03 to 1.73 and the transparency was good.
The telescope was strongly defocused to average pixel-to-pixel sensitivity 
variations and to increase the duty cycle, resulting in a typical full width at 
half-maximum of the stellar images of $\sim$9 pixels ($\sim$5.8\arcsec). The 
integration time was 6 s.  After a standard pre-reduction (bias, dark and 
flat-field correction), the stellar fluxes were extracted from the images using 
the {\tt IRAF/DAOPHOT}\footnote{{\tt IRAF} is distributed by the National 
Optical Astronomy Observatory, which is operated by the Association of 
Universities for Research in Astronomy, Inc., under cooperative agreement with 
the National Science Foundation.} aperture photometry software 
\citep{1987PASP...99..191S}.  We tested several sets of reduction parameters and 
chose the set that gave the most precise photometry for the stars of similar 
brightness to WASP-45. We carefully selected a set of reference stars and then 
performed differential photometry.

The WASP-45b transit light curves are shown in Figure~\ref{fig:w45-phot} and the 
data are given in Table~\ref{tab:phot}.

\subsubsection{WASP-46b}

A transit of WASP-46b was observed on 2010 July 19 by TRAPPIST. 
Observations lasted from 1h15 to 4h05 UT, during which the airmass decreased 
from 1.83 to 1.22 and the transparency was good. Considering the relative 
faintness of the target, we chose to keep the telescope focused during the run, 
and we used an integration time of 60 s. The pre-reduction and reduction 
procedures were similar to those used for the WASP-45 data. 

A transit of WASP-46b was observed on 2010 September 10 by Euler. We 
observed for a total of 3.4 hours, from 35 min before the start of transit 
to 75 min after it ended. The conditions were variable and airmass increased 
from 1.12 to 1.4 during the observations. No defocus was applied.

The WASP-46b transit light curves are shown in Figure~\ref{fig:w46-phot} and the 
data are given in Table~\ref{tab:phot}.

\begin{table} 
\caption{Euler and TRAPPIST photometry of the three stars} 
\label{tab:phot} 
\begin{tabular*}{0.5\textwidth}{@{\extracolsep{\fill}}llllll} 
\hline 
Set & Star & Tel. & BJD(UTC) & Rel.flux & $\Delta$Rel.flux\\ 
& & & $-2\,450\,000$ & & \\
\hline
1	& WASP-44 & Euler & 5453.686865 & 0.9989 & 0.0015 \\
1	& WASP-44 & Euler & 5453.687942 & 1.0007 & 0.0015 \\
1	& WASP-44 & Euler & 5453.689683 & 1.0000 & 0.0015 \\
\ldots	& \ldots & \ldots & \ldots & \ldots & \ldots \\
5	& WASP-46 & Euler & 5449.614796 & 0.9991 & 0.0014 \\
5	& WASP-46 & Euler & 5449.616092 & 0.9978 & 0.0014 \\
5	& WASP-46 & Euler & 5449.617396 & 1.0003 & 0.0014 \\
\hline
\end{tabular*}
This table is available in its entirety in the online journal. 
\end{table}

\section{Stellar parameters from spectra}
\label{sec:stellar-params}
For each system, the CORALIE spectra were co-added to produce a
single spectrum with a typical S/N of around 50:1. 
The analysis was performed using the methods given in 
\citet{2009A&A...501..785G}.
The \halpha\ line was used to determine the
effective temperature (\teff), while the Na {\sc i} D and Mg {\sc i} b lines
were used as surface gravity (\logg) diagnostics. The parameters obtained from
the analysis are given in Table~\ref{tab:stellar}. 
The elemental abundances were determined from equivalent-width measurements of 
several clean and unblended lines. A value for microturbulence (\mictrb) was 
determined from Fe~{\sc i} using the method of \citet{1984A&A...134..189M}. The 
quoted error estimates include that given by the uncertainties in \teff, \logg\ 
and \mictrb, as well as the scatter due to measurement and atomic data 
uncertainties. 

\begin{table*}
\caption{Stellar parameters from spectra}
\begin{tabular}{lccc}
\hline
Parameter (Unit) & WASP-44 & WASP-45 & WASP-46 \\
\hline
\teff\ (K)	& $5400 \pm 150$ & $5100 \pm 200$ & $5600 \pm 150$\\
\logg\ (cgs)	& $4.5 \pm 0.2$  & $4.4 \pm 0.2$  & $4.4 \pm 0.2$\\
\mictrb\ (\kms)	& $1.0 \pm 0.2$    & $0.5 \pm 0.3$   & $1.0 \pm 0.2$\\
\vsini\ (\kms)	& $3.2 \pm 0.9$    & $2.3 \pm 0.7$   & $1.9 \pm 1.2$\\
{[Fe/H]} & $0.06 \pm 0.10$ & $0.36 \pm 0.12$ & $-0.37 \pm 0.13$\\
{[Na/H]} & $0.24 \pm 0.09$ & $0.56 \pm 0.14$ & $-0.35 \pm 0.09$\\
{[Mg/H]} & $0.25 \pm 0.07$ & $0.57 \pm 0.09$ & $-0.10 \pm 0.10$\\
{[Al/H]} & $0.20 \pm 0.13$ & $0.45 \pm 0.11$ & $-0.24 \pm 0.08$\\
{[Si/H]} & $0.25 \pm 0.13$ & $0.50 \pm 0.15$ & $-0.27 \pm 0.10$\\
{[Ca/H]} & $0.10 \pm 0.17$ & $0.38 \pm 0.17$ & $-0.21 \pm 0.13$\\
{[Sc/H]} & $0.30 \pm 0.17$ & $0.45 \pm 0.27$ & $-0.13 \pm 0.16$\\
{[Ti/H]} & $0.12 \pm 0.10$ & $0.46 \pm 0.20$ & $-0.29 \pm 0.18$\\
{[V/H]}  & $0.24 \pm 0.17$ & $0.73 \pm 0.18$ & $-0.21 \pm 0.21$\\
{[Cr/H]} & $0.13 \pm 0.07$ & $0.34 \pm 0.12$ & $-0.31 \pm 0.12$\\
{[Mn/H]} & $0.29 \pm 0.10$ & $0.93 \pm 0.24$ & $-0.47 \pm 0.14$\\
{[Co/H]} & $0.23 \pm 0.09$ & $0.70 \pm 0.11$ & $-0.32 \pm 0.24$\\
{[Ni/H]} & $0.13 \pm 0.12$ & $0.51 \pm 0.16$ & $-0.39 \pm 0.10$\\
$\log A$(Li)$^{\rm a}$ & $<1.0$          & $<0.8$          & $<0.8$\\
\mstar\ (\msol) & $0.95 \pm 0.08$ & $0.95 \pm 0.10$ & $0.93 \pm 0.09$\\
\rstar\ (\rsol) & $0.90 \pm 0.22$ & $1.00 \pm 0.25$ & $1.00 \pm 0.26$\\
Spec. Type      & G8V             & K2V             & G6V \\
Gyro. Age & $0.9^{+1.0}_{-0.6}$ & $1.4^{+2.0}_{-1.0}$ & $1.4^{+0.4}_{-0.6}$\\
\hline
R.A. (J2000)	& $\rm  00^{h} 15^{m} 36\fs76$       & $\rm  00^{h} 20^{m} 56\fs99$       & $\rm  21^{h} 14^{m} 56\fs86$ \\
Dec. (J2000)	& $\rm -11\degr 56\arcmin 17\farcs4$ & $\rm -35\degr 59\arcmin 53\farcs8$ & $\rm -55\degr 52\arcmin 18\farcs1$ \\
$m_{\rm V}$		& $12.9$		& $12.0 \pm 0.2$	& $12.9$\\
$m_{\rm J}$$^{\rm b}$	& $11.70 \pm 0.02$	& $10.75 \pm 0.02$	& $11.75 \pm 0.02$\\
$m_{\rm H}$$^{\rm b}$	& $11.41 \pm 0.03$	& $10.37 \pm 0.03$	& $11.47 \pm 0.03$\\
$m_{\rm K}$$^{\rm b}$	& $11.34 \pm 0.03$	& $10.29 \pm 0.02$	& $11.40 \pm 0.03$\\
2MASS$^{\rm b}$		& 00153675$-$1156172	& 00205699$-$3559537	& 21145687$-$5552184\\
\hline
\end{tabular}
\label{tab:stellar}
\\
$^{\rm a}$ $\log A$(Li)  = $\log N($Li/H$) + 12$, where $N$(Li/H]) is the number 
density of lithium with respect to hydrogen.\\
$^{\rm b}$ \citet{2006AJ....131.1163S}.
\end{table*}

The projected stellar rotation velocities (\vsini) were determined by fitting 
the profiles of several unblended Fe~{\sc i} lines. 
For this we used an instrumental FWHM of 0.11 $\pm$ 0.01 \AA\ (= 5.2 $\pm$ 0.5 
\kms), determined from the telluric lines around 6300\AA\ (equating to a 
spectral resolution of $\sim$57\,000), and used the \cite{2010MNRAS.405.1907B} 
calibration to assume values for macroturbulence (\mactrb). The \mactrb\ values 
we assumed were 1.4 $\pm$ 0.3 \kms\ for WASP-44, 0.7 $\pm$ 0.3 \kms\ for 
WASP-45, and 2.0 $\pm$ 0.3 \kms\ for WASP-46. 

Though the S/N are very low in the Ca {\sc ii} H+K regions of our spectra, we 
searched for signs of emission, which would be indicative of stellar 
chromospheric 
activity. The WASP-44 spectra show no signs of emission and both the WASP-45 and 
WASP-46 spectra show weak emission. 

We input our values of \teff, \logg\ and \feh\ into the calibrations of 
\citet{2010A&ARv..18...67T} to obtain estimates of the stellar mass and radius. 

We used the method of \citet{2007ApJ...669.1167B} to estimate gyrochronological 
ages for the stars. 
For WASP-44 and WASP-45 we calculated rotation periods from our \vsini\ 
determinations. If either star is inclined with respect to the sky plane then it 
will be rotating faster than suggested by \vsini\ and so is likely to be 
younger than indicated.
For WASP-46, the rotation period found from the modulation of the WASP light 
curves was used (see Section~\ref{sec:rot}). As such, the age estimate for this 
star is not just an upper limit.

\section{System parameters from RV and transit data}
\label{sec:syspar}
\subsection{WASP-44b and WASP-46b}
We determined the parameters of each system from a simultaneous fit to all data. 
The fit was performed using the current version of the 
Markov-chain Monte Carlo (MCMC) code described by \citet{2007MNRAS.380.1230C} 
and \citet{2008MNRAS.385.1576P}. 

The transit light curves were modelled using the formulation of 
\citet{2002ApJ...580L.171M} with the assumption that the planet is much smaller 
than the star. 
Limb-darkening was accounted for using a four-coefficient, nonlinear 
limb-darkening model, using coefficients appropriate to the passbands from the 
tabulations of \citet{2000A&A...363.1081C, 2004A&A...428.1001C}. 
The coefficients were interpolated once using the values of \logg\ and \feh\ in 
Table~\ref{tab:stellar}, but were interpolated at each MCMC step using the 
latest value of \teff. The coefficient values corresponding to the best-fitting 
value of \teff\ are given in Table~\ref{tab:ld}.
The transit light curve is parameterised by the epoch of mid-transit 
$T_{\rm 0}$, the orbital period $P$, the planet-to-star area ratio 
(\rplanet/\rstar)$^2$, the approximate duration of the transit from initial to 
final contact $T_{\rm 14}$, and the impact parameter $b = a \cos i/R_{\rm *}$ 
(the distance, in fractional stellar radii, of the transit chord from the 
star's centre in the case of a circular orbit), where $a$ is the semimajor axis 
and $i$ is the inclination of the orbital plane with respect to the sky plane. 

\begin{table*}
\caption{Limb-darkening coefficients} 
\label{tab:ld} 
\begin{tabular}{lccccccc}
\hline
Planet		& Instrument	& Observation bands		& Claret band	& $a_1$		& $a_2$		& $a_3$		& $a_4$	\\
\hline
WASP-44		& WASP / Euler	& Broad (400--700 nm) / Gunn $r$& Cousins $R$	& 0.646 	& $-$0.437 	& 1.080		& $-$0.519 \\
WASP-45		& WASP / Euler	& Broad (400--700 nm) / Gunn $r$& Cousins $R$	& 0.755 	& $-$0.768 	& 1.450		& $-$0.623 \\
WASP-45		& TRAPPIST	& Cousins $I$+Sloan $z'$	& Sloan $z'$	& 0.833		& $-$0.891 	& 1.304		& $-$0.547 \\
WASP-46		& WASP / Euler	& Broad (400--700 nm) / Gunn $r$& Cousins $R$	& 0.543 	& $-$0.055 	& 0.592		& $-$0.348 \\
WASP-46		& TRAPPIST	& Cousins $I$+Sloan $z'$	& Sloan $z'$	& 0.630		& $-$0.371 	& 0.727		& $-$0.366 \\
\hline
\end{tabular}
\end{table*}

The eccentric Keplerian radial-velocity orbit is parameterised by the stellar 
reflex velocity semi-amplitude $K_{\rm 1}$, the systemic velocity $\gamma$, and 
\secos\ and \sesin\ (Collier Cameron, in preparation),  where $e$ is orbital 
eccentricity and $\omega$ is the argument of periastron. 

The linear scale of the system depends on the orbital separation $a$ which, 
through Kepler's third law, depends on the stellar mass \mstar. 
At each step in the Markov chain, the latest values of \densstar, \teff\ and 
\feh\ are input in to the empirical mass calibration of 
\citet{2010A&A...516A..33E} to obtain \mstar.
The shapes of the transit light curves and the radial-velocity curve constrain 
\densstar\ \citep{2003ApJ...585.1038S}, which combines with \mstar\ to give 
\rstar.
\teff\ and \feh\ are proposal parameters constrained by Gaussian priors with 
mean values and variances derived directly from the stellar spectra 
(see Section~\ref{sec:stellar-params}). 

As the planet-star area ratio is determined from the measured transit depth, 
\rplanet\ follows from \rstar. The planet mass \mplanet\ is calculated from the 
the measured value of $K_{\rm 1}$ and the value of \mstar; the planetary density 
\densplanet\ and surface gravity $\log g_{\rm pl}$ then follow. 
We  calculate the planetary equilibrium temperature \teql, assuming zero 
albedo and efficient redistribution of heat from the planet's 
presumed permanent day side to its night side. 
We also calculate the durations of transit ingress ($T_{\rm 12}$) and egress 
($T_{\rm 34}$), the total occultation duration ($T_{\rm 58}$), and 
the durations of occultation ingress ($T_{\rm 56}$) and egress ($T_{\rm 78}$).

At each step in the MCMC procedure, model transit light curves and radial 
velocity curves are computed from the proposal parameter values, which are 
perturbed from the previous values by a small, random amount. The \chisq\ 
statistic is used to judge the goodness of fit of these models to the data and a 
step is accepted if \chisq\ is lower than for the previous step. A step 
with higher \chisq\ is accepted with a probability 
proportional to $\exp(-\Delta \chi^2/2)$, 
which gives the procedure some robustness against local minima and leads to the 
thorough exploration of the parameter space around the best-fitting solution. 
To give proper weighting to each photometry data set, the uncertainties were 
scaled at the start of the MCMC so as to obtain a photometric reduced-\chisq\ 
of unity. 
For the same reason, a jitter term of 10.2 \ms\ was added in quadrature to the 
formal errors of the WASP-46 RVs. 
A possible source of this jitter is stellar activity, as indicated by 
chromospheric and photometric indicators (Sections~\ref{sec:stellar-params} and 
\ref{sec:rot}).
With reduced spectroscopic \chisq\ values of less than unity, it was not 
necessary to add any jitter to the RVs of WASP-44.

\subsection{WASP-45b}
The analysis for WASP-45b was the same as for WASP-44b and WASP-46b, except we 
used evolutionary models to impose priors on stellar mass and density for 
reasons to be explained.

The apparent lack of second and third contact points in the transit light 
curves of WASP-45b (Figure~\ref{fig:w45-phot}) indicate that the transit is 
grazing or near-grazing. 
When b + \rplanet/\rstar $>$ 1, the planet does not fully pass in front of 
the star. 
Thus, as $b$ tends toward higher values, \rplanet\ must 
inflate so that the ratio \rplanet/\rstar\ remains consistent with the observed 
transit depth. Also, \rstar\ must inflate so as to remain consistent with the 
observed transit duration. 
This results in stellar and planetary radii that are non-physical in a sizeable 
portion of accepted MCMC steps. 

\begin{figure}
\centering                     
\includegraphics[width=8.4cm]{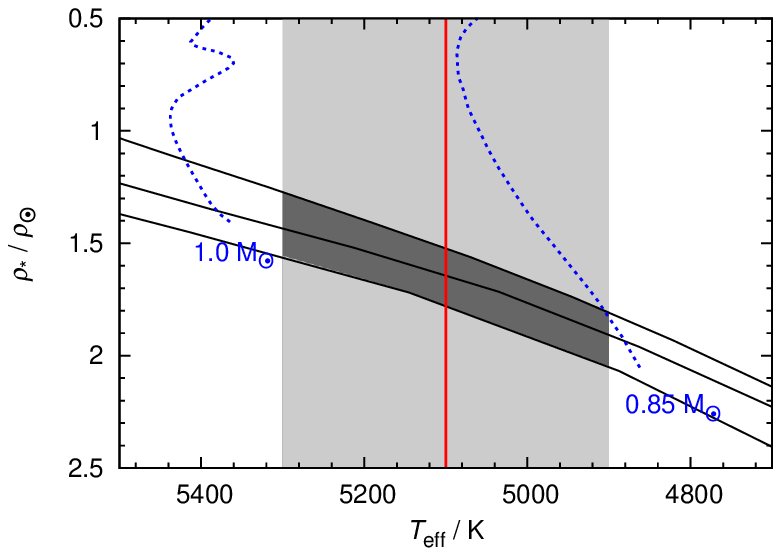}
\caption{Evolutionary diagram for WASP-45 with the models of 
\citet{2004ApJS..155..667D}. 
The solid, black lines are isochrones, with the ages coming from gyrochronology: 
\feh\ = 0.48, 3.4 Gyr (top); 
\feh\ = 0.36, 1.4 Gyr (middle); 
\feh\ = 0.24, 0.4 Gyr (bottom). 
The two dashed, blue lines are the mass tracks for 1.0 \msol\ with \feh\ = 0.48 
and 0.85 \msol\ with \feh\ = 0.24. 
The isochrones and mass tracks we chose to display are those that bound the 
region of interest, which is the area shaded dark grey that indicates the 
range of stellar densities consistent with the models. 
The solid, red line indicates our best-fitting \teff, 
and the area shaded light grey corresponds to the 1-$\sigma$ confidence 
interval.
}
\label{fig:w45-evol}
\end{figure}

To avoid biasing the derived stellar and planetary radii to larger values, 
we imposed a prior on stellar density using the constraints provided by 
our gyrochronological age determination (Table~\ref{tab:stellar}) and 
evolutionary models. 
Our gyrochronolgical age is determined using \vsini\ and so is an upper limit. 
If the stellar spin axis is inclined with respect 
to the sky plane then the star will truly be rotating faster and will therefore 
be younger. 
We plotted the isochrones and mass tracks of \citet{2004ApJS..155..667D} 
relevant to the value ranges we obtained for \teff, \feh\ and age 
(Figure~\ref{fig:w45-evol}). 
By interpolating the models for these measured values, we obtained stellar 
dimensions consistent with the models: \densstar\ = $1.65 \pm 0.40$ \denssol\ 
and \mstar\ = $0.91 \pm 0.06$ \msol. 
We used these values to place Gaussian priors, by means of Bayesian penalties on  
\chisq, at each MCMC step.

It was not necessary to add any jitter to the RVs of WASP-45.

\subsection{The imposition of circular orbits}
For each system, the best-fitting eccentricity is small and 
consistent with zero (Table~\ref{tab:ecc}). 
We used the $F$-test approach of \citet{1971AJ.....76..544L} to calculate the 
probability, \PLS, that the improvement in the fit that results 
from fitting for an eccentric orbit could have arisen by 
chance if the underlying orbit were circular. 
For each system, we found \PLS\ to be much higher (Table~\ref{tab:ecc}) than 
the threshold of 0.05 suggested by \citet{1971AJ.....76..544L}, below which they 
consider the detection of a non-zero eccentricity to be significant. 
In the absence of conclusive evidence to the contrary, we assumed the orbit of 
each planet to be circular in producing the adopted solutions that we present. 
This issue is dealt with at greater length in the discussion.

\begin{table*} 
\caption{The limits placed on orbital eccentricity and the impact of fitting for it} 
\label{tab:ecc}
\begin{tabular}{lccc}
\hline
Parameter (Unit) & WASP-44 & WASP-45 & WASP-46\\ 
\hline 
\\
rms$_{\rm spec}$ (\ms) & 24.8 & 8.8 & 33.9\\
$\chi^2_{\rm spec}$ \medskip & 10.5 & 6.3 & 18.1\\
$e$ & 0.036$^{+ 0.054}_{- 0.025}$& 0.023$^{+ 0.027}_{- 0.016}$& 0.018$^{+ 0.021}_{- 0.013}$\\
\PLS & 0.51 & 1 & 0.67\\
$e$ (2 $\sigma$ upper limit) & $<0.172$ & $<0.095$ & $<0.065$\\
$e$ (3 $\sigma$ upper limit) & $<0.263$ & $<0.148$ & $<0.092$\\
$\omega$ ($^\circ$) & $-59^{+ 132}_{- 47}$& $21^{+59}_{-94}$& 65.$^{+ 50}_{- 122}$\\
$e\cos\omega$ & 0.003$^{+ 0.023}_{- 0.018}$& 0.010$^{+ 0.015}_{- 0.011}$& 0.001$^{+ 0.014}_{- 0.010}$\\
$e\sin\omega$ & $-0.011^{+ 0.028}_{- 0.069}$& 0.003$^{+ 0.034}_{- 0.019}$& 0.008$^{+ 0.026}_{- 0.012}$\\
$\phi_{\rm mid-occultation}$ & 0.502$^{+ 0.015}_{- 0.012}$& 0.5061$^{+ 0.0096}_{- 0.0067}$& 0.5007$^{+ 0.0087}_{- 0.0066}$\\
$T_{\rm 58}$ (d) & 0.0915$^{+ 0.0049}_{- 0.0093}$& $0.0711^{+0.0055}_{-0.0073}$& 0.07008$^{+ 0.00089}_{- 0.00104}$\\
$T_{\rm 56} \approx T_{\rm 78}$ (d) \medskip & 0.0132$^{+ 0.0033}_{- 0.0030}$& $>0.0257$ & 0.0184$^{+ 0.0029}_{- 0.0018}$\\
$M_{\rm *}$ ($M_{\rm \odot}$) & $0.948 \pm 0.034$ & $0.910 \pm 0.060$ & $0.957 \pm 0.034$\\
$R_{\rm *}$ ($R_{\rm \odot}$) & 0.892$^{+ 0.085}_{- 0.097}$& $0.950^{+0.093}_{-0.074}$ & $0.928 \pm 0.034$\\
$\rho_{\rm *}$ ($\rho_{\rm \odot}$) \medskip & 1.33$^{+ 0.54}_{- 0.30}$& $1.06 \pm 0.27$& $1.20 \pm 0.12$\\
\mplanet\ (\mjup) & 0.893$^{+ 0.071}_{- 0.066}$& $1.005 \pm 0.053$ & $2.100 \pm 0.073$\\
\rplanet\ (\rjup) & $1.09^{+ 0.13}_{- 0.14}$ & 1.17$^{+0.28}_{-0.14}$ & $1.327 \pm 0.058$\\
\densplanet\ (\densjup) & 0.69$^{+ 0.37}_{- 0.20}$& $0.64 \pm 0.30$& $0.90 \pm 0.12$\\
\\ 
\hline 
\end{tabular}
\\
This table demonstrates the effects of fitting for orbital eccentricity and is 
to be compared with Table~\ref{tab:mcmc}, which presents our adopted solutions.
\end{table*}

\subsection{Results}
The median values and 1-$\sigma$ uncertainties of the system parameters derived 
from the MCMC model fits are presented in Table~\ref{tab:mcmc}. 
The corresponding transit models are superimposed on the transit photometry in 
Figures~\ref{fig:w44-phot}, \ref{fig:w45-phot} and \ref{fig:w46-phot}. 
The corresponding orbit models are superimposed on the radial velocities in 
Figures~\ref{fig:w44-rv}, \ref{fig:w45-rv} and \ref{fig:w46-rv}, in which we 
also show the RV residuals as a function of time. From the residual plots one 
can evaluate both the level to which stellar activity may have affected the RVs 
and the evidence for a third body. 
The RVs of each planet-hosting star, obtained over the course of $~$100 
days, show no evidence for motion induced by a third body. 
The residuals of WASP-46 show greater scatter (\chisq\ = 19.2 for WASP-46 
compared to 10.9 for WASP-44 and 7.4 for WASP-45), which is probably due to the 
star being active (see Sections~\ref{sec:stellar-params} and \ref{sec:rot}). 

\begin{table*} 
\caption{System parameters from RV and transit data from our adopted, circular 
solutions} 
\label{tab:mcmc}
\begin{tabular}{lccc}
\hline
Parameter (Unit) & WASP-44 & WASP-45 & WASP-46\\ 
\hline 
\\
$P$ (d) & $2.4238039 \pm 0.0000087$ & $3.1260876 \pm 0.0000035$ & $1.4303700 \pm 0.0000023$\\
$T_{\rm 0}$ (HJD, UTC) & $2\,455\,434.376 \pm 0.00040$ & $2\,455\,441.26925 \pm 0.00058$ & $2\,455\,392.31553 \pm 0.00020$\\
$T_{\rm 14}$ (d) & $0.0936 \pm 0.0022$ & $0.0746 \pm 0.0035$ & $0.06973 \pm 0.00090$\\
$T_{\rm 12} = T_{\rm 34}$ (d) & $0.0146 \pm 0.0026$& $>0.0270$ & $0.0178 \pm 0.0013$\\
$a/\rstar$ & $8.05^{+0.66}_{-0.52}$ & $9.22 \pm 0.74$ & $5.74 \pm 0.15$\\
\rplanet$^2$/\rstar$^2$ & $0.01588 \pm 0.00076$ & $0.01495 \pm 0.00071$ & $0.02155 \pm 0.00049$\\
$b$ & 0.560$^{+ 0.076}_{- 0.123}$& $0.888^{+0.045}_{-0.026}$& $0.737 \pm 0.019$\\
$i$ ($^\circ$) \medskip & 86.02$^{+ 1.11}_{- 0.86}$ & $84.47^{+0.54}_{-0.79}$& $82.63 \pm 0.38$\\
$K_{\rm 1}$ (\ms) & $138.8 \pm 9.0$ & $148.3 \pm 4.1$ & $387 \pm 10$\\
$\gamma$ (\ms) & $-4\,043.82 \pm 0.71$ & $4\,549.95 \pm 0.63$ & $-3\,778.3 \pm 1.1$\\
$e$ & 0 (adopted) & 0 (adopted) & 0 (adopted) \\
rms$_{\rm spec}$ (\ms) & 25.0 & 9.7 & 34.2\\
$\chi^2_{\rm spec}$ \medskip & 10.9 & 7.4 & 19.2\\
$M_{\rm *}$ ($M_{\rm \odot}$) & $0.951 \pm 0.034$ & $0.909 \pm 0.060$ & $0.956 \pm 0.034$\\
$R_{\rm *}$ ($R_{\rm \odot}$) & $0.927^{+0.068}_{-0.074}$ & $0.945^{+0.087}_{-0.071}$& $0.917 \pm 0.028$\\
$\log g_{*}$ (cgs) & 4.481$^{+ 0.068}_{- 0.057}$& $4.445^{+0.065}_{-0.075}$ & $4.493 \pm 0.023$\\
$\rho_{\rm *}$ ($\rho_{\rm \odot}$) & 1.19$^{+ 0.32}_{- 0.22}$& $1.08^{+0.27}_{-0.24}$& $1.24 \pm 0.10$\\
$T_{\rm eff}$ (K) & $5410 \pm 150$ & $5140 \pm 200$ & $5620 \pm 160$\\
{[Fe/H]} \medskip & $0.06 \pm 0.10$ & --- & $-0.37 \pm 0.13$\\
\mplanet\ (\mjup) & $0.889 \pm 0.062$& $1.007 \pm 0.053$ & $2.101 \pm 0.073$\\
\rplanet\ (\rjup) & $1.14 \pm 0.11$ & $1.16^{+0.28}_{-0.14}$ & $1.310 \pm 0.051$\\
$\log g_{\rm pl}$ (cgs) & 3.197$^{+ 0.094}_{- 0.082}$& $3.23^{+0.11}_{-0.19}$& $3.447 \pm 0.033$\\
\densplanet\ (\densjup) & 0.61$^{+ 0.23}_{- 0.15}$& $0.64 \pm 0.30$ & $0.94 \pm 0.11$\\
$a$ (AU)  & $0.03473 \pm 0.00041$ & $0.04054 \pm 0.00090$ & $0.02448 \pm 0.00028$\\
$T_{\rm pl, A=0}$ (K) & $1343 \pm 64$ & $1198 \pm 69$ & $1654 \pm 50$\\
\\ 
\hline 
\end{tabular} 
\end{table*} 

We also performed MCMCs fitting an eccentric model to the RVs so as to 
illustrate the impact on those parameters that are liable to be affected when 
imposing a circular orbit. 
The best-fitting values and associated uncertainties from MCMCs fitting an 
eccentric model are given in Table~\ref{tab:ecc}, and these are to be compared 
with the values of our adopted solutions in Table~\ref{tab:mcmc}, derived using 
a circular model. 
For each system, the best-fitting eccentricity is small and the improvement 
in the fit is slight. 
The parameter values of the circular and eccentric models agree to well within 
1 $\sigma$, and the associated uncertainties of the eccentric model are only 
fractionally larger.

\section{Rotational modulation}
\label{sec:rot}
\begin{figure*} 
\begin{center}
\includegraphics[width=0.9\textwidth]{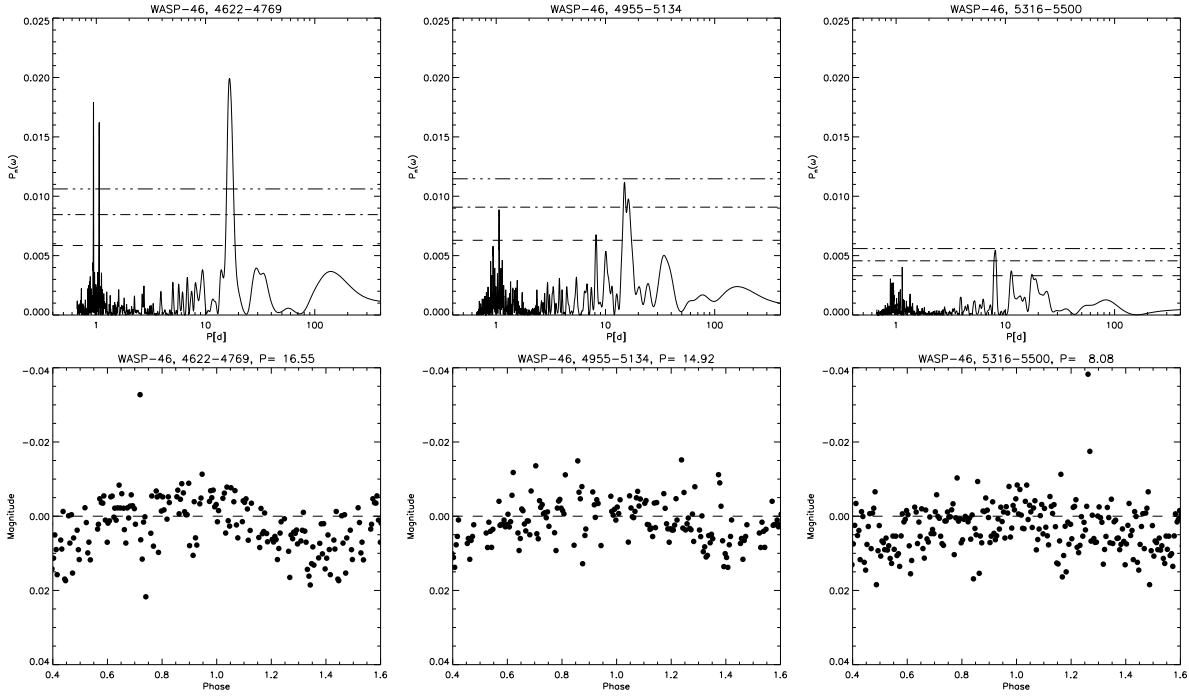}
\end{center}
\caption{{\it Upper panels}: Periodograms for the WASP data from three seasons 
for WASP46. The date range (JD$-$245000) is given in the title of each panel.
Horizontal lines indicate false alarm probability levels FAP=0.1, 0.01, 0.001. 
{\it Lower panels}: Data  folded and binned in 0.005 phase bins for the periods
noted alongside the date range in the plot title.}
\label{fig:swlomb}
\end{figure*}

We analysed the WASP light curves of each star to determine whether they show 
periodic modulation due to the combination of magnetic activity and stellar 
rotation.  
We used the sine-wave fitting method described in \citet{2011PASP..123..547M} to 
calculate periodograms such as those shown in the upper panels of 
Figure~\ref{fig:swlomb}.
These are calculated over 4\,096 uniformly spaced frequencies from 0 to 1.5
cycles day$^{-1}$. The false alarm probability levels shown in these figures are
calculated using a boot-strap Monte Carlo method also described in 
\citet{2011PASP..123..547M}. Variability due to star spots is not expected to be 
coherent on long timescales as a consequence of the finite lifetime of 
star-spots and differential rotation in the photosphere so we analysed each  
season of data separately. 

We found no evidence of rotational modulation in any of the WASP-44 or 
WASP-45 light curves. 

We applied the sine-wave fitting method to the WASP-46 light curves from each of 
the three seasons to produce the periodograms in the upper panels of 
Figure~\ref{fig:swlomb}.
The peaks in the periodograms for the first two seasons of data are highly
significant and  imply  periods of  $16.55 \pm 0.10$\,d and $14.92 \pm
0.07$\,d. We used a boot-strap Monte Carlo method to estimate the errors on
these periods. The periodogram for the third season of data shows peaks 
corresponding to a period of 8.08\,d. Our boot-strap Monte Carlo method
fails to provide a reliable error estimate for this less significant peak. The
amplitudes of the variations estimated from the sine-wave fits are in the range 
of 3--5 milli-magnitudes (Fig.~\ref{fig:swlomb}, lower panels). 
We also calculated periodograms of two stars with similar
magnitudes and colours to WASP-46 observed simultaneously with the same
camera. Neither of these stars showed significant periodic modulation in the
same frequency interval with the exception of one season of data for one star
that showed significant power close to 1 cycle day$^{-1}$, the source of which 
is likely to be diurnal.  

Our interpretation of these results is that the WASP light curve of WASP-46
does show periodic modulation due to the rotation of the star with a period of
$16\pm 1$\,days. The period of 8.08\,days derived from the third season of
data can be explained by the distribution of star spots during this season
resulting in a light curve with two minima per cycle. Similarly, the poorly
defined peak in the periodogram for the second season of data may be due to
the appearance and disappearance of star spot groups during the observing
season.

Using the rotational-modulation period of $16 \pm 1$\,days and the stellar 
radius from Table~\ref{tab:mcmc}, we calculate the stellar rotation velocity 
$v$ to be $2.9 \pm 0.2$ \kms. 
That this value is higher than the value of \vsini\ ($1.9 \pm 1.2$ \kms) 
determined from spectral line profiles (Section~\ref{sec:stellar-params}) 
suggests that the stellar spin axis is inclined by $I = 41^{+40}_{-27}$ $^\circ$ 
with respect to the sky plane. This should be appreciated when assessing whether 
a measurement of WASP-46b's Rossiter-McLaughlin effect is indicative of an 
aligned or a misaligned orbit \citep[e.g.][]{2010ApJ...719..602S}. 
Usually only the sky-projected angle between the planetary orbit and 
stellar spin axis can be determined, but with a measurement of $I$ the true 
angle can be determined. 
Our measurement of $I$ is imprecise and would be improved by a more precise 
determination of spectroscopic \vsini. 

\section{Discussion}

WASP-44b is a 0.89-\mjup\ planet in a 2.42-day orbit around a solar metallicity 
(\feh\ = $0.06 \pm 0.10$) G8V star. The star is observable from both 
hemispheres and seems to be inactive as it shows no signs of rotational 
modulation or Ca {\sc ii} H+K emission. 

WASP-45b is a Jupiter-mass planet that partially transits its metal rich (\feh\ 
= $0.36 \pm 0.12$) K2V host star every 3.13 days. 
The (near-)grazing nature of the transits led us to impose a prior on the 
density of the host star so that the derived stellar and planetary radii are 
physical. 
The star may be active as weak Ca {\sc ii} H+K emission is seen in its spectra, 
though no rotational modulation is evident in the light curves.

WASP-46b is a 2.10-\mjup\ planet in a 1.43-day orbit around a metal poor (\feh\ 
= $-0.37 \pm 0.13$) G6V star. 
We found the star to be active as weak Ca {\sc ii} H+K emission is present in its 
spectra and a clear 16-day rotational modulation signature is evident in the 
WASP light curves. 
The slower rotation rate suggested by spectrocopic \vsini\ indicates that the 
stellar spin axis is inclined relative to the sky plane 
($I = 41 \pm 40 ^\circ$). 
By observing WASP-46b's Rossiter-McLaughlin effect 
\citep[e.g.][]{2011MNRAS.414.3023S}, which is predicted to have 
an amplitude of 18 \ms, we could thus measure the true angle between the 
stellar spin and the planetary orbit axes, rather than just the sky-projected 
angle.
The large size of the planet, the high stellar insolation, and the 
relative lateness of the host star make WASP-46b a good target for infrared 
occultation photometry \citep[e.g.][]{2011MNRAS.416.2108A} . 

The upper limits placed on the lithium abundances suggest that each star is at 
least a few Gyr old \citep[Table~\ref{tab:stellar};][]{2005A&A...442..615S}, 
which is in contrast with the gyrochronological ages of $\sim$0.9--1.4 Gyr.

\subsection{Fit for eccentricity or impose a circular orbit?}

To obtain accurate stellar densities and the derivative stellar and 
planetary radii, we require not only high-quality transit light curves, but also 
accurate determinations of the orbital eccentricities. 
Measurements of these parameters inform intense theoretical efforts concerning 
tidal circularisation and heating, bulk planetary composition and the 
observed systematic errors in planetary and stellar radii. 
For those planets with poorly constrained eccentricity, further radial-velocity 
and occultation-timing measurements are necessary. 
However, these will be performed slowly and perhaps only for a fraction of 
systems. 
As such, we should make every effort to avoid introducing biases into the 
sample that we have to work with. 

Around half of all known transiting planets (i.e. those confirmed by a dynamical 
mass determination) are hot Jupiters 
($P \lesssim 4$ d and \mplanet\ $\approx$ 0.5--2 \mjup) and 
\citet{2011MNRAS.tmp..378P} find no evidence for an eccentric orbit among these. 
Further, tidal theory predicts that the timescale of tidal circularisation 
is a sharp function of semimajor axis and planet radius, and a weaker function 
of planetary and stellar masses \citep{1966Icar....5..375G}. 
That is, low-mass, bloated planets in short orbits around high-mass stars are 
expected to circularise quickest.  
So, there is a theoretical basis and an empirical basis for expecting a 
newly-discovered hot Jupiter to be in a circular orbit. 

When a hot Jupiter is first announced the radial-velocity data often poorly 
constrain eccentricity and occultation observations are almost never available. 
In those circumstances, some authors have opted to impose a circular orbit 
(e.g. \citealp{2011A&A...531A..60A}, \citealp{2011A&A...531A..40F}), 
as is theoretically and empirically reasonable, though doing so  
does artificially reduce the uncertainties on the system paramters. 
Other authors have opted to fit for an eccentric orbit 
(e.g. \citealp{2010ApJ...723L..60H}, \citealp{2011ApJ...742..116B}) so as to 
obtain more conservative error bars. 
Fitting an eccentric orbit model to RV measurements of a 
(near-)circular orbit will always result in a non-zero value of $e$ 
\citep[e.g.][]{2005ApJ...629L.121L}, and this can affect the derived stellar 
and planetary parameters. 
As compared to the circular solution, the eccentric solution will typically 
change the planet's speed during transit and, as the transit duration is fixed 
by measurement, a change in speed necessitates a change in distance travelled 
during transit, i.e. the derived stellar radius. In turn, as the ratio of 
planetary-to-stellar radii is fixed by the measured transit depth, then the 
derived planet radius is changed. 

Being aware of these issues, some authors have opted to present both a circular 
solution, which is most likely, and an eccentric solution, which gives a 
reliable account of the uncertainties in the data 
\citep[e.g.][]{2010ApJ...709..159A}. 
One issue with this approach is that many authors tend to prefer to work with a 
single solution and so will adopt the one they favour regardless. 
Another approach is to present the best-fitting parameter values from a circular 
fit and the error bars from an eccentric fit 
(e.g. \citealp{2011MNRAS.410.1631E}, \citealp{2011PASP..123..547M}). 
Thus the most likely solution is presented with conservative error bars, but 
care must be taken to calculate the 1-$\sigma$ confidence intervals in the 
posterior probability distributions about the best-fitting circular values. 

As the best-fitting eccentricity values for the three systems presented in this 
paper are small ($e$ = 0.02--0.04) the differences between the best-fitting 
parameter values from circular and from eccentric solutions are small 
(compare Tables~\ref{tab:ecc} and \ref{tab:mcmc}). 
However, eccentricity is often poorly determined in a discovery paper and the 
choice to fit for eccentricity then has a greater impact. 
An illustrative example is the case of WASP-17b. In the discovery paper 
\citep{2010ApJ...709..159A}, we presented three solutions, including one with an 
imposed circular orbit which gave \rplanet\ = $1.97 \pm 0.10$ \rjup, and 
another, which we adopted as our preferred solution, in which we fit for an 
eccentric orbit that gave \rplanet\ = $1.74^{+0.26}_{-0.23}$ \rjup\ and 
$e =0.129^{+0.106}_{-0.068}$. 
We had initially favoured the circular solution as theory suggested this as the 
most likely and the additional two free parameters of an eccentric orbit did 
not significantly improve the fit. 
However, we adopted the eccentric solution due to its more conservative error 
bars. 
We had been able to place only a weak constraint on eccentricity as the low mass 
of the planet (0.49 \mjup) and the high effective surface temperature of the 
star (6650 K) caused our measurement of the reflex motion of the star to be 
quite low S/N.
This situation improved when we observed two occultations of WASP-17b, which we 
used to show the orbit of WASP-17b to be very slightly eccentric 
\citep[$e \approx 0.0055$][]{2011MNRAS.416.2108A} and the planet radius to be 
$1.991 \pm 0.081$ \rjup. As such, the circular solution that we presented in the 
discovery paper was closest to describing the system and we would have done 
better to have adopted that. 

By the time we had shown the orbit of WASP-17b to be near-circular, a number of 
papers had, at least in part, based their models, 
results or conclusions on the values of one or both of $e$ and \rplanet\ of the 
discovery paper's adopted solution 
(e.g. \citealp{2010ApJ...719..602S}, \citealp{2010ApJ...723..285H}, 
\citealp{2011ApJ...726...82C}, \citealp{2011A&A...529A.136E}). 
Though it is true that our updated solution was not excluded, not even at the 
1-$\sigma$ level, some of those studies had only considered WASP-17 due to its 
apparent significant eccentricity, and other such works were sure to follow. 
For example, the possibility that tidal dissipation could have inflated WASP-17b 
would probably have been tested as was done by \citet{2011ApJ...727...75I} 
for other systems. 
Before we showed the planet to be even more bloated than first suggested 
and placed a much tighter constraint on eccentricity, tidal heating had seemed 
a much more likely explanation for the anomalously large size of WASP-17b.

A more recent example, which is similar to the case of WASP-17 and became 
available after submission of this paper, is that of HAT-P-32 and HAT-P-33 
\citep{2011ApJ...742...59H}. 
In both systems, the signal-to-noise of the stellar radial motion, as induced by 
the orbiting planet, is low and eccentricity is poorly constrained. 
In an early draft (arXiv:1106.1212v1), eccentric solutions were presented 
alongside circular solutions such that the reader could choose which to adopt. 
The derived planetary radii differ significantly between the eccentric and the 
circular solutions, with 1.8 \rjup\ versus 2.0 \rjup\ for HAT-P-32b and 
1.7 \rjup\ versus 1.8 \rjup\ for HAT-P-33b. 
In the published version of the paper, \citet{2011ApJ...742...59H} instead 
suggest adoption of the circular solutions, but caution that several eccentric 
short-period planets are known. 
The three example exoplanets given, XO-3b, WASP-14b and HAT-P-21b, are each 
much more massive and smaller than HAT-P-32b and HAT-P-33b, and are in similar 
or longer orbits. 
Each effect acts to increase the tidal circularisation time-scale. Thus, it is 
little wonder that these planets are in eccentric orbits, but it would be 
suprising if HAT-P-32b or HAT-P-33b were. 

Unless there is clear evidence for an eccentric orbit, we recommend imposing a 
circular orbit when deriving the parameters of a hot-Jupiter system. 
This will avoid creep in the literature of parameter values as further RV and 
occultation observations show discovery-paper eccentricities to be spurious. 
Though further measurements will show some planets to have a small, non-zero 
eccentricity, for the majority eccentricity will be constrained to ever-smaller 
values consistent with zero. 

\section*{Acknowledgements}
WASP-South is hosted by the South African Astronomical Observatory and  
SuperWASP-N is hosted by the Issac Newton Group on La Palma. We are 
grateful for their ongoing support and assistance. Funding for WASP comes from 
consortium universities and from the UK's Science and Technology Facilities 
Council.
TRAPPIST is a project funded by the Belgian Fund for Scientific Research (Fond 
National de la Recherche Scientifique, FNRS) under the grant FRFC 2.5.594.09.F, 
with the participation of the Swiss National Science Fundation (SNF).  M. 
Gillon and E. Jehin are FNRS Research Associates.


\label{lastpage}

\end{document}